\documentstyle[multicol,prl,aps,epsf]{revtex} 

\begin{document}


\title{{\it History} Memorized and Recalled upon Glass Transition}
\author{Yoshihisa Miyamoto, Koji Fukao,}
\address{Faculty of Integrated Human
 Studies, Kyoto University, Kyoto 606-8501 Japan}  
 \author{Hiromi Yamao,}
\address{Graduate School of Human and Environmental Studies, 
Kyoto University, Kyoto 606-8501 Japan}  
\author{and Ken Sekimoto}
\address{Yukawa Institute of Theoretical Physics, 
Kyoto University, Kyoto 606-8502 Japan\footnote{
Present address: UMR 168 - Institut Curie, 26, rue d'Ulm, 
75248 Paris Cedex 05, France}
}  

\maketitle

\begin{abstract}
The memory effect upon glassification is studied in
the glass to rubber transition of vulcanized rubber
with the strain as a controlling parameter.
A phenomenological model is proposed
taking the history of the temperature and the strain into account, 
by which the experimental results are interpreted.
The data and the model demonstrate that the glassy state memorizes the 
time-course of strain upon glassification, not as a single parameter 
but as the history itself.
The data also show that the effect of irreversible deformation
in the glassy state is beyond the scope of the present model. \\
Authors' remark: 
The title of the paper in the accepted version is above.
The title appeared in PRL is the one changed by a Senior Assistant
Editor after acceptance of the paper.
The recovery of the title was rejected in the correction process.
\end{abstract}

\begin{multicols}{2}

There are a group of phenomena where 
the exchange of the relative magnitude of the system's relaxation time
and the timescale of observation plays an important role. 
They are the subject of active interest of both 
non-equilibrium statistical physics
and material sciences\cite{Bellon99,Djurberg99}.
Examples are the glass transition, fluidization, plasticization, etc.,
for which the relaxation time depends on the temperature, the
concentration of solute or plasticizer, 
the applied force and so on\cite{Derec00,Derec01,Cloitre00,Leibler93}. 
This paper addresses a general problem for such phenomena:
In what manner and to what extent memories can be stored upon freezing 
as its internal frozen state, and also can be read out upon defreezing?

In order to study the memory effects in the frozen state,
it is versatile to introduce 
an experimentally controllable parameter other than the temperature
to control the relaxation time of the system.
In the present paper, 
the following methods are adopted for this purpose:
(1) We establish the experimental method to measure
the glass to rubber transition of rubber under strain. 
We prepare samples in glassy states by three different protocols, and 
study the glass to rubber transition of those samples using 
two different modes of measurement.
(2) Also we propose a phenomenological model equation as a natural 
extension of the linear viscoelastic model.
We show both experimentally and theoretically that the time-course of 
the strain itself is memorized as a history in the glassy state. 
(3) We show the limitation of the present model to cover 
the irreversible properties in the glassy state.

The material studied here is made of synthetic {\it cis}-1,4 polyisoprene
kindly supplied by Toyo Tire \& Rubber Co., Ltd.
The number of monomer units between crosslinks is estimated to be
190 from the Mooney-Rivlin elastic coefficient, and
200 from the equilibrium swelling ratio by the use of toluene as a solvent
where the polymer-solvent interaction parameter
of 0.34 is employed\cite{Treloar75,PolymerHB89}.
The samples, typically 20mm$\times$2mm$\times$0.5mm in their original size,
are placed in a temperature controlled
(less than $\pm 1^{\circ}$C above $-90^{\circ}$C)
ethanol bath, except for the case of stretching at room temperature,
where it is done in the air.
The length of sample $\ell$ is monitored by a displacement sensor
controlled by a motor and the tension $f$ is measured by a force gauge.
The nominal stress is given by $\frac{f}{d_0 w_0}$
and the strain, in terms of the elongation ratio
$\lambda = \frac{\ell}{\ell_0} = 1 + \gamma$,
where $\ell_0$, $d_0$ and $w_0$ are the original length, thickness
and width, respectively.
No correction is made for the thermal expansion of the rubber.

Three protocols to prepare the stretched glassy states are the following.
{\it (HT)} The samples are stretched at room temperature
(the rubber region) followed by quenching to $-55^{\circ}$C and cooled to
$-100^{\circ}$C at ca. 3K/min.
{\it (GT)} The samples are quenched to $-61 \pm 1^{\circ}$C
(the glass transition region), stretched
at a strain rate, $\dot{\gamma}$, of 0.045 s$^{-1}$
and kept at a prescribed 
length for a given time, and then cooled to $-100^{\circ}$C.
{\it (LT)} The samples are quenched to $-76$ or $-90 \pm 1^{\circ}$C
(the glassy state), then stretched and cooled to $-100^{\circ}$C.
In the last protocol
a neck is formed and propagated on stretching
at a draw speed of 0.18 mm/s.

Two modes of measurement of the glass to rubber transition are employed:
{\it Stress-free mode} -- 
After releasing the stress of  stretched glasses at $-100^{\circ}$C,
the length of the sample is measured on heating at 1K/min
under the stress-free condition.
{\it Isometric mode} -- 
At $-100^{\circ}$C, the stress is released 
(except for one case, see below), and then the length of the sample is fixed
(Upon release of the stress, the elongation ratio
changes  typically from  4.03 to 3.99$_5$). 
After that the tension is measured on heating at 1K/min
with the length being kept fixed. 

\begin{figure}[htbp]
  \epsfxsize=8cm
  \begin{center}
  \epsffile{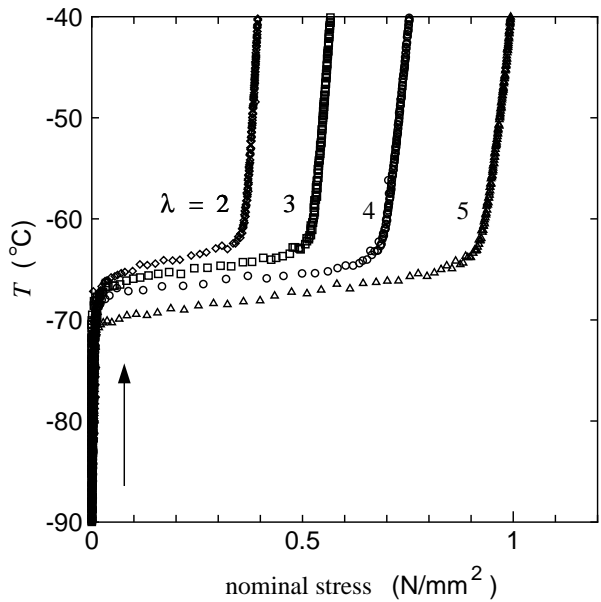}
  \end{center}
  \caption{Stress recovery on heating at 1K/min under the fixed-length 
condition (isometric mode) for the rubbers stretched at room temperature 
(protocol-HT). The elongation ratios at room temperature are 
2.04, 3.04, 4.03 and 5.07 from the left to the right.
}
  \label{fig:rtrlsh1}
\end{figure}

Figure \ref{fig:rtrlsh1} shows the stress recovery
upon the glass to rubber transition in the isometric mode
of the sample prepared by the protocol-HT.
No stress emerges in the glassy state
and the stress recovers the rubbery values corresponding 
to a given strain at the glass-rubber transition.
The glass transition temperature $T_{\rm g}$ is determined as
the center temperature of the stress recovery process; 
$T_{\rm g}$ decreases with the elongation ratio
(see also Fig.\ref{fig:TgSL} below).
In the rubber region, the tension is approximately proportional
to the absolute temperature,
and the decrease in tension arising from crystallization, which is
observed only for $\lambda=5$
above $-35^{\circ}$C, is outside of the range of the data shown.

\begin{figure}[htbp]
  \epsfxsize=8cm
  \begin{center}
  \epsffile{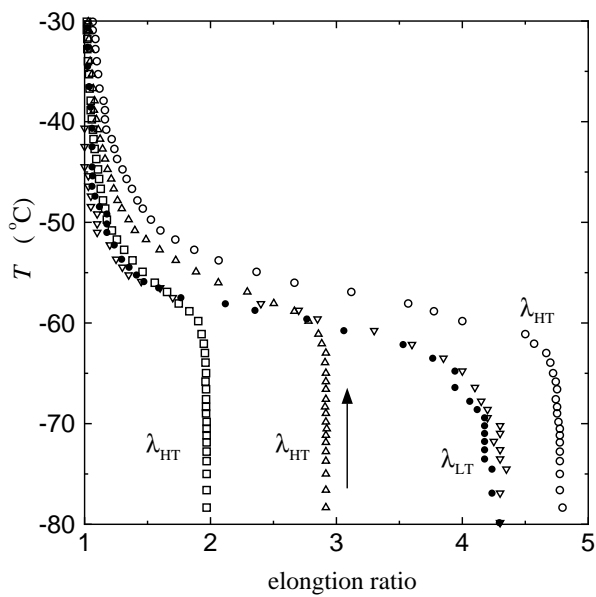}
  \end{center}
  \caption{Contraction of glassified rubber on heating at 1K/min 
under the unloaded condition (stress-free mode). 
$\Box$, $\triangle$ and $\circ$, the samples stretched at room
temperature (protocol-HT). $\bullet$, stretched at $-76^{\circ}$C and
$\bigtriangledown$, $-90^{\circ}$C (protocol-LT).
}
  \label{fig:LT1}
\end{figure}

Figure \ref{fig:LT1} shows the variation in elongation ratio
of the sample with temperature under the stress-free mode.
The samples are graduated in ink in 0.5mm intervals at room temperature,
and we monitor the change in the length by a video recorder.
In the glassy state, the samples maintain their length,
and start shrinking at the glass-rubber transition down to the original length.
The shrinkage of the graduation on rubber is macroscopically homogeneous
in both protocols-HT and -LT.
The bending and coiling on contraction can be avoided by
placing the sample in transparent plastic tubes with holes.
For the samples prepared by the protocol-HT
($\lambda_{\rm HT}$),
the temperature to start shrinking is about $-60^{\circ}$C
and decreases slightly with elongation.
The neck part of the samples prepared by the protocol-LT
has similar elongation ratio, $\lambda_{\rm LT} \sim 4$,
and this part shrinks macroscopically homogeneously. 
The shrinking starts, however, at lower temperatures than the samples
prepared in the protocol-HT by about $5^{\circ}$C.

The data by the isometric-mode and stress-free modes
(Figs.\ref{fig:rtrlsh1} and \ref{fig:LT1}) show
the similar effects of elongation on the glass to rubber transition
for the samples prepared by the protocol-HT, but
the glass to rubber transition is observed at lower temperatures
in Fig.\ref{fig:rtrlsh1} than Fig.\ref{fig:LT1} by several degrees.
The difference can be ascribed to the method of measurement:
the characteristic times observed by the strain
and by the stress measurements differ by the ratio of
the relaxed modulus to the unrelaxed one\cite{McCrum67}.

\begin{figure}[htbp]
  \epsfxsize=8cm
  \begin{center}
  \epsffile{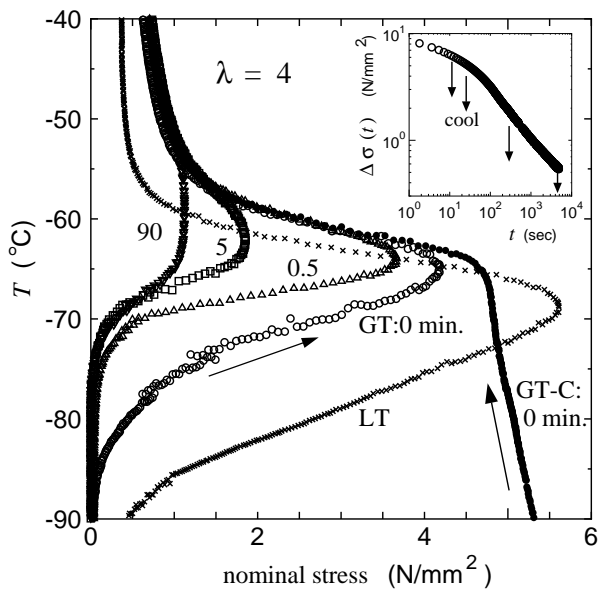}
  \end{center}
  \caption{
Stress recovery on heating in the isometric mode.
The elongation ratio is about 4. The samples are stretched 
at $-61^{\circ}$C and the stress is relaxed for:
$\circ$, 10 s, $\triangle$, 0.5 min, $\Box$, 5 min, 
$\bigtriangledown$, 90 min (protocol-GT)
and $\bullet$, 10 s (GT-C: protocol-GT and isometric mode without
releasing stress at $-100^{\circ}$C). $\times$,
The sample stretched at $-76^{\circ}$C (protocol-LT). 
Note the difference of the scale from Fig.\ref{fig:rtrlsh1}.
Inset: The aging of the stress at $T=-61^{\circ}$C after
keeping at $\lambda=4$. $\Delta \sigma (t) = \sigma (t) - \sigma(\infty)$, 
where $\sigma(\infty)$ is determined by the value at $-61^{\circ}$C for
$\lambda = 4$ in Fig.\ref{fig:rtrlsh1}.
}
  \label{fig:rlsh1}
\end{figure}

The inset of Fig.\ref{fig:rlsh1} shows the aging of stress at $-61^{\circ}$C 
in the protocol-GT, just after the stretching process.
After four different aging times indicated by the arrows,
the sample is cooled down to $-100^{\circ}$C.
The results of stress recovery on heating in the isometric mode 
are shown in the main part of Fig.\ref{fig:rlsh1},
together with a result for a sample prepared by the protocol-LT.
For comparison we also show a datum ($\bullet$, denoted by GT-C)
of the sample whose history is the same as the one denoted
by GT:0 min($\circ$), except that the  %
stress was not released at $-100^{\circ}$C.
In the last sample, the tension decreases slowly with
temperature below $-65^{\circ}$C because, at least partly, of the 
thermal expansion of the sample.
In the remaining samples, on the other hand,  
the tension shows a prominent maximum where
the temperature is nearly the $T_{\rm g}$ of the sample of the protocol-HT
with $\lambda_{\rm HT}=4$.
The effect of aging is evident:
the temperature at which the tension departs from zero increases
as a function of the aging time.
The sample prepared by the protocol-LT
shows a still more pronounced memory effect.

The dependence of $T_{\rm g}$ on elongation ratio is shown
in Fig.\ref{fig:TgSL}. For the samples prepared by the protocol-HT,
$T_{\rm g}$ decreases with strain and stress.
On the other hand, the samples prepared by the protocol-GT,
whose $T_{\rm g}$ is defined here as the peak temperature in 
Fig.\ref{fig:rlsh1}, show various values of $T_{\rm g}$ 
in spite of nearly the same elongation ratios. 

We propose the following phenomenological model equation
for the stress $\sigma(t)$ at the time $t$ as a functional of the 
strain $\gamma(t')$ and the temperature $T(t')$.
\begin{eqnarray}
\sigma(t)\!\! = \!\sigma_{\rm R} (T(t),\! \gamma(t))\!\! +\!
G_\infty \!\!\! \int^{t}_{-\infty} \!\!\!\!\!\!\! 
[{\gamma}(t)\!\!-\!{\gamma}(t')] 
 \frac{\partial {\cal G}\!\left( \tilde{t}(t,t')\right) }{\partial t'}
{\rm d}t'\!,\!\!\!\!
\label{eq:ve1}
\end{eqnarray}
where $\sigma_{\rm R}$ represents a rubber-like response,
$G_\infty$ is the modulus of the glass well below $T_{\rm g}$.
In our experiment the Young's modulus of the stretched glass 
at $-100^{\circ}$C is about $2 \times 10^9 \; {\rm Pa}$ 
assuming 0.5 for the Poisson's ratio,
while the rubber-like
Young's modulus at room temperature is $8.6 \times 10^5 \; {\rm Pa}$.
$\tilde{t}(t,t')$ is defined by $\tilde{t}(t,t')\equiv
\int^t_{t'} [\tau(T(u),\gamma (u))]^{-1} {\rm d}u$,
giving the {\it intrinsic} time lapse between $t'$ and $t(>t')$,
which is measured with the instantaneous relaxation time $\tau(T,\gamma)$ 
as a function of $T$ and $\gamma$. 
${\cal G}(z)$ is the scaled relaxation function which decreases from 
${\cal G}(0)=1$ to ${\cal G}(\infty)=0$.
Upon integration by parts, 
Eq.(\ref{eq:ve1}) reduces to the form similar to 
the familiar linear viscoelasticity equation\cite{Bird87},
$\int^{t}_{-\infty} {\cal G}\left(\tilde{t}(t,t')\right) 
\dot{\gamma}(t'){\rm d}t'$,
but the original form Eq.(\ref{eq:ve1}) is physically more inspiring
for the present purpose: 
The second term on the right hand side of Eq.(\ref{eq:ve1}) reads
that, the fraction 
$\frac{\partial}{\partial t'}{\cal G}\left(\tilde{t}(t,t')\right)  {\rm d}t'$
of the mechanical elements has become effective since the time
slice $[t',t'+dt']$, and each engaged element 
contributes to the stress by
$G_\infty [{\gamma}(t)-{\gamma}(t')]$ at $t$.
In the special case of ${\cal G}(z)=e^{-z}$, 
Eq.(\ref{eq:ve1}) reduces to the Maxwell-type model,
$(\frac{{\rm d}}{{\rm d}t}+\frac{1}{\tau(\gamma,T)})[\sigma-\sigma_{\rm R}]
=G_\infty \frac{{\rm d} \gamma}{{\rm d}t}$,
which has been proposed independently
of us in \cite{Derec01} to study the aging, where
the temperature dependence of $\tau$ was not scrutinized. 

Below, the results of our experiment
are explained qualitatively
on the basis of Eq.(\ref{eq:ve1}), except for the protocol-LT.
We assume that for $T>T_{\rm g}(\gamma)$ the relaxation time 
$\tau(T,\gamma)$ is of the microscopic order while 
for $T<T_{\rm g}(\gamma)$ it is beyond the experimental timescale.
If we stretch the rubber at $T>T_{\rm g}(\gamma)$, the integral of
Eq.(\ref{eq:ve1}) is $\sim \dot{\gamma}\tau\ll 1$, and the stress
is rubber-like: $\sigma \simeq \sigma_{\rm R}$.
If the temperature is decreased rapidly across $T_{\rm g}(\gamma)$
within a short time, ${\cal G}\left(\tilde{t}(t,t')\right)$ increases suddenly
from $\simeq 0$ to $\simeq 1$, because $\tilde{t}(t,t')$ decreases 
abruptly. Thus the integral in Eq.(\ref{eq:ve1})
will yield $\simeq \gamma(t)-\gamma(t_0)$,
where $t_0$ is the time at which $T$ reaches well below 
$T_{\rm g}(\gamma)$, say, $T = -100^\circ$C.
As long as the sample is in the glassy state, $\tilde{t}$ does not
increase, and the system behaves as a solid elastic body.
At this stage, the stress is released (except for the protocol GT-C
mentioned above), say at $t_1$($>t_0$).
Thus $\gamma_1\equiv \gamma(t_1)$ satisfies
\begin{equation}
0=\!\sigma_{\rm R}(T,\!\gamma_1)\!\! +\! G_\infty \!\!\!
\int^{t_0}_{-\infty}  \!\!\!\!\!\!\!  [\gamma_1\!\! -\!\gamma(t')]
\frac{\partial}{\partial t'}{\cal G}\!(\tilde{t}(t_0,t'))dt'\!. \!\!\!\!
\label{eq:vef}
\end{equation}
As long as the temperature is well below $T_{\rm g}(\gamma(t_0))$,
$\tilde{t}(t,t')=\tilde{t}(t_0,t')$ holds in the 
integral of Eq.(\ref{eq:ve1}), thus $t_1$ does not appear in this expression.
If we fix $\gamma(t)$ at $\gamma_1$ upon reheating (isometric mode),
$\sigma(t)$ of Eq.(\ref{eq:ve1}) departs from zero when $T(t)$ approaches
$T_{\rm g}$ from below:  $\tilde{t}(t,t')$ abruptly increases
and the integral tends towards 0 rapidly, leaving only the rubber-like
elasticity for $\sigma(t)$ (Fig.\ref{fig:rtrlsh1}).
On the other hand, the reheating under the stress-free mode
leads to a funny phenomenon:
In response to the decrease of the integral in Eq.(\ref{eq:ve1}),
the rubber-like tension $\sigma_{\rm R}$ tends to shrink the sample.
However, the decrease of $\gamma(t)$ is {\it decelerated} when the point 
$T(t)=T_g(\gamma(t))$ is approached.
As the result, the strain $\gamma(t)$ nearly traces the part of the curve
of $T=T_g(\gamma)$, as the temperature is gradually 
raised (Fig.\ref{fig:LT1}).
Note that the situation would be completely different for the systems 
with  ${\rm d}T_g/{\rm d}\gamma>0$. 

For the sample prepared by the protocol-GT, the integrand
$[\gamma_1-\gamma(t')]\frac{\partial}{\partial t'}{\cal G}$
in Eq.(\ref{eq:vef}) 
is non-negligible for a substantial range of $t'$ up to $t_0$,
since $\tau$ is comparable to the experimental timescale in this case.
When we start reheating, say after $t=t_2$,
we have $\tilde{t}(t,t')=\tilde{t}(t_0,t')+\tilde{t}(t,t_2)$.
Since $\tilde{t}(t,t_2)$ is independent of $t'$, this
process leads to a common ``shift''
of the function ${\cal G}$ by  $\alpha$, 
${\cal G}(z)\mapsto {\cal G}(z+\alpha)$,
in the integral of Eq.(\ref{eq:ve1}),
with  $\alpha =\tilde{t}(t,t_2)$. 
Except for the special case of ${\cal G}(z+\alpha)\propto {\cal G}(z)$
for all $z$  (i.e. ${\cal G}=e^{-z}$), 
the equality (\ref{eq:vef}) will be broken
under the above shift of ${\cal G}$ at $t>t_2$,
implying that the 
nonzero stress reappears upon reheating under the isometric mode
(Fig.\ref{fig:rlsh1}).
This is the origin of the memory effect.
A numerical check has been done with 
${\cal G}(z)=
\theta {\rm e}^{-\frac{z}{z_1}}
+ (1-\theta) {\rm e}^{-\frac{z}{z_2}}$
and $\tau(T,\gamma_2)$ at the glass to rubber 
transition region being a constant independent of $T(t)$. 
The result (not shown) of $\sigma$ as 
the function of the shift $\alpha\equiv \tilde{t}(t,t_2)$
reproduces the memory effect. Its peak amplitude
decreases with the duration of the aging, while the peak
position is found at virtually the same value for various values of $\alpha$.
More extensive results will be discussed elsewhere.

A critically important remark is that, in the memory effect,
the stress $\sigma(t)$ depends on the history of 
$\gamma(t')$ and $T(t')$ before $t_0$ {\it not} through a
single ``order parameter''.
It is true even if $\tau$ was constant at around $T_{\rm g}$: 
we may regard the integral of Eq.(\ref{eq:ve1}) as the linear
transformation with the ``matrix'' $\cal M$
whose $(z,\alpha)$ component is
${\cal G}(z+\alpha)$ with $z\ge 0$ and $\alpha \ge 0$.
The transformation can then preserve certain amount of information of
the ``vector'' $\gamma(t')$ with
$t'\le t_0$ according to the rank of the matrix ${\cal M}$, i.e.,
the number of independent ``columns'' of the matrix $\cal M$. If, for example,
${\cal G}(z)$ consists of $N$ exponentially decaying functions,
the rank is $N$.
Thus, it is the time course of the control parameters
that is memorized and remembered upon the glass transition.
The relation between the aging and the memory effect 
should be discussed in future in this context.

In summary, 
the glass-rubber transition was studied for the stretched glasses
with the strain as a controlling parameter.
The results including the memory effect are interpreted by
a phenomenological formula.
The form of Eq.(\ref{eq:ve1}) is very general, and it may also apply, 
for example, to dipole glasses where the relaxation may obey the similar
relaxational law. The present framework, however, 
cannot capture the prominent memory effect
of the sample stretched in the glassy state (protocol-LT),
whose aspect will be addressed elsewhere.

The authors would like to thank F. Lequeux for helpful discussions
and for introducing refs.\cite{Bellon99,Djurberg99},
and R. Ohara for supplying the material.
This work is partly supported by a Grant-in-Aid of
Japan Society for the Promotion of Science and the Yamada Science 
Foundation.

\begin{figure}[htbp]
  \epsfxsize=8cm
  \begin{center}
  \epsffile{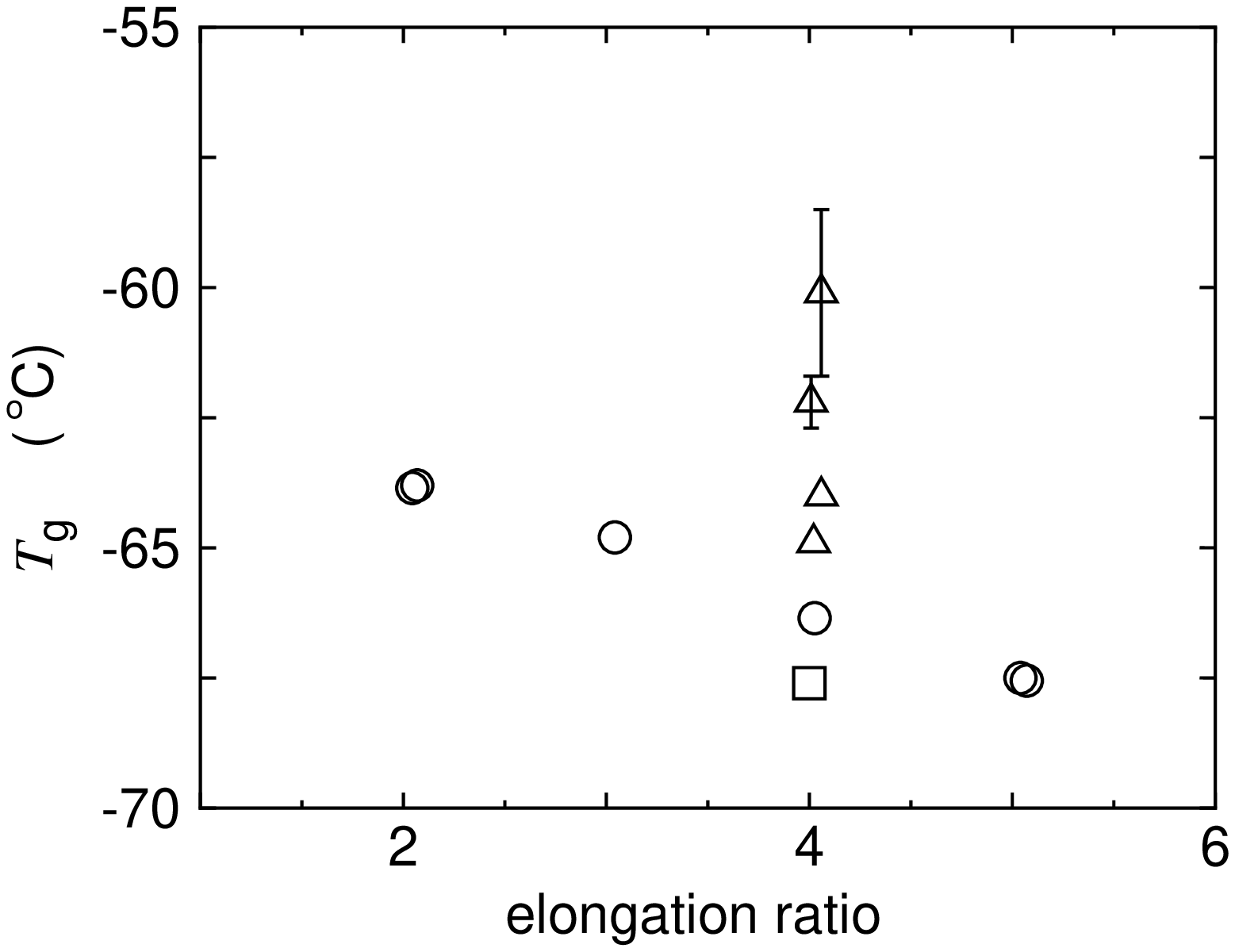} 
  \end{center}
  \caption{Dependence of $T_{\rm g}$ on elongation ratio
(isometric mode): $\circ$,
protocol-HT, $\triangle$, protocol-GT and $\Box$, protocol-LT.             
The values of $T_{\rm g}$ prepared in protocol-GT decrease
as function of the stress measured at $-100^{\circ}$C (not shown).
}
  \label{fig:TgSL}
\end{figure}

\end{multicols}


\begin{references}
\bibitem{Bellon99}
L. Bellon, S Ciliberto and C. Laroche,
cond-mat/9906162.
\bibitem{Djurberg99}
C. Djurberg, K. Jonason and P. Nordblad,
Euro. Phys. J., {\bf B10}, 15 (1999).
\bibitem{Derec00}
C. Derec, A. Ajdari, G. Ducouret and F. Lequeux, 
C. R. Acad. Sci. Paris t. 1, S\'{e}rie IV, 1115 (2000).
\bibitem{Derec01}
C. Derec, A. Ajdari and F. Lequeux, 
Eur. Phys. J., E{\bf 4}, 355 (2001).
See, especially, Eq.(4a) there.
\bibitem{Cloitre00}
M.Cloitre, R. Borrega and L. Leibler, 
Phys. Rev. Lett., {\bf 85}, 4819 (2000).
\bibitem{Leibler93}
L. Leibler and K. Sekimoto, 
Macromolecules, {\bf 26}, 6937 (1993).
\bibitem{Bird87}
R. B. Bird, R. C. Armstrong and O. Hassager,
{\it  Dynamics of Polymeric Liquids}, vol.1
John Wiley \& Sons, (1987) Chapter 5.
\bibitem{Treloar75}
L. R. G. Treloar, 
{\it The Physics of Rubber Elasticity}, 3rd. ed., 
Clarendon Press, Oxford, (1975) Chapters 5 and 7.
\bibitem{PolymerHB89}
{\it Polymer Handbook}, 3rd. ed., 
ed. by J. Brandup and E. H. Immergut, 
A Wiley International Publication, (1989) Chapter 7.
\bibitem{McCrum67}
N. G. McCrum, B. E. Read and G. Williams,
{\it Anelastic and Dielectric Effects in Polymer Solids},
John Wiley \& Sons Ltd, (1967) Chapter 4.
\end{references}
\end{document}